\documentstyle[aps,prl,manuscript,graphicx]{revtex}
\begin{document}
\author{A. M. C. Souza$^{1, \,2}$ and C. Tsallis$^1$ \thanks{%
amcsouza@ufs.br, tsallis@cbpf.br}}
\address{$^1$Centro Brasileiro de Pesquisas Fisicas \\
Rua Xavier \\
Sigaud 150, 22290-180 Rio de Janeiro-RJ, Brazil\\
$^2$Departamento de Fisica \\
Universidade Federal de \\
Sergipe, 49100-000 Sao Cristovao-SE, Brazil }
\title{Stability of the entropy for superstatistics}
\maketitle

\begin{abstract}
The Boltzmann-Gibbs celebrated entropy $S_{BG}=-k\sum_ip_i \ln p_i$ is {\it %
concave} (with regard to all probability distributions $\{p_i\}$) and {\it %
stable} (under arbitrarily small deformations of any given probability
distribution). It seems reasonable to consider these two properties as {\it %
necessary} for an entropic form to be a {\it physical} one in the
thermostatistical sense. Most known entropic forms (e.g., Renyi entropy)
violate these conditions, in contrast with the basis of nonextensive
statistical mechanics, namely $S_q=k\frac{1-\sum_ip_i^q}{q-1} \;(q\in {\cal R%
}; \;S_1=S_{BG})$, which satisfies both ($\forall q>0$). We have recently
generalized $S_q$ (into $S$) in order to yield, through optimization, the
Beck-Cohen superstatistics. We show here that $S$ satisfies both conditions
as well. Given the fact that the (experimentally observed) optimizing
distributions are invariant through {\it any} monotonic function of the entropic form to be optimized, this might constitute a very strong criterion for
identifying the physically correct entropy.
\end{abstract}

%\section{ Introduction}

\vspace{1in}

The Boltzmann-Gibbs (BG) entropy expression 
\begin{equation}
S_{BG} =-k\sum_{i=1}^W p_i \ln p_i \;,
\end{equation}
(sometimes referred to as the {\it Boltzmann-Gibbs-Shannon entropy}), and
its equal-probability particular case $S_{BG}=k \ln W$, undoubtedly
constitute a major step in human understanding of the underlying laws of
nature. Its continuous form $S_{BG} =-k \int dx p(x) \ln p(x)$, historically
introduced by Boltzmann and Gibbs, its quantum form $S_{BG}=-k\, Tr \, \rho
\ln \rho$, introduced by von Neumann, and the above mentioned discrete form
(1), introduced by Shannon (though in a different context), have been at the
basis of invaluable scientific results in physics, chemistry and elsewhere.
Let us learn more from it. The optimization of Eq. (1) under the canonical
ensemble (equilibrium with a thermostat) constraints $\sum_{i=1}^Wp_i=1$ and 
$\sum_{i=1}^W p_i E_i=U_{BG}$ ($U_{BG}$ is the internal energy, and $\{E_i\}$
are the eigenvalues of the Hamiltonian), yields the celebrated BG
probability for the equilibrium stationary state 
\begin{equation}
p_i \propto e^{-\beta E_i} \;\;\;(\beta \equiv 1/kT)
\end{equation}
This distribution can always be, at least in principle, experimentally
checked (and it has been so with well known success). Not so $S_{BG}$!
Although it normally goes without discussion, it is important to notice, for
the arguments we want to develop here, that {\it any} monotonic function of $%
S_{BG}$ (e.g., $S_{BG}^{\, 3}/k^2$) would provide, through optimization, the
very same distribution as Eq. (2). BG statistical mechanics is however much
more than just the equilibrium distribution, and there are plethoric reasons
for considering the {\it physical entropy} to be precisely $S_{BG}$ and no
other function of it. We will quickly realize that the situation can become
much less trivial as soon as we want to generalize BG statistical mechanics,
in order to address other (typically nonergodic) stationary states such as
those which ubiquitously emerge in complex systems.

In order to handle nonequilibrium states of large classes of systems
(examples nowadays known include turbulence \cite
{turbulencebeck1,turbulencebeck2,turbulencearimitsu}, electron-positron
annihilation \cite{bediaga}, quark-gluon plasma \cite{rafelski}, anomalously
diffusing micro-organisms \cite{hydra}, classical \cite{classicalchaos} and
quantum \cite{quantumchaos} chaos, long-range many-body Hamiltonians \cite
{longrange}, economics \cite{lisa}, and others; see \cite{reviews} for
reviews), one of us postulated \cite{tsallis} the entropic form 
\begin{equation}
S_q=k\frac{1-\sum_{i=1}^W p_i^q}{q-1} \;\;\;(q\in {\cal R}; \;S_1=S_{BG}) \;,
\end{equation}
as the basis for generalizing BG statistical mechanics. It can be trivially
verified that $p_{ij}^{A+B}=p_i^A \times p_j^B$, $\forall (i,j)$, implies $%
S_q(A+B)/k=S_q(A)/k+S_q(B)/k+(1-q)S_q(A)S_q(B)/k^2$, property which makes
the generalized formalism to be sometimes referred to as {\it nonextensive
statistical mechanics}. If we optimize $S_q$ with the already mentioned norm
constraint and $\sum_{i=1}^Wp_i^qE_i/ \sum_{i=1}^Wp_i^q=U_q$, we obtain \cite
{TMP} 
\begin{equation}
p_i \propto e_q^{-\beta_q^\prime E_i} \;,
\end{equation}
where the $q$-exponential function is defined as $e_q^x \equiv
[1+(1-q)x]^{1/(1-q)}\;(e_1^x=e^x)$, and $\beta_q^\prime$ is an effective
inverse temperature. This distribution is ubiquitously detected in nature.
The question arises: {\it is $S_q$ the proper physical entropy for
generalizing the BG formalism?} Indeed, any monotonic function of $S_q$,
such as for example the Renyi entropy $S_q^R\equiv [\ln
\sum_{i=1}^Wp_i^q]/(1-q)=(1-q)^{-1}\ln[1+(1-q)S_q]$ (we use $k=1$ from now
on), will also yield Eq. (4) if optimized under the same constraints. The
answer seems to be {\it yes}. Indeed, $S_q$ is, $\forall q>0$, both {\it %
concave} (with regard to all probability distributions $\{p_i\}$) and {\it %
stable} (under arbitrarily small deformations of any given probability
distribution), whereas $S_q^R$ is {\it not} (like all other known entropic
forms which are monotonic functions of $S_q$) \cite{lesche,abestable}.

The entropy $S_{BG}$ can be obtained from appropriate set of axioms, as long
shown by Shannon and also by Khinchin. However, the same occurs with $S_q$
and $S_q^R$ (and others) by modifying in admissible manners Shannon' s
and/or Khinchin's sets of axioms \cite{santosabe}. Therefore this path does
not help much for distinguishing the physical entropy from its monotonic
functions. But concavity and stability do. Indeed, no monotonic function of $%
S_q$ is known which satisfies both. On top of this, $S_q$ allows, in all
known cases (both vanishing and nonvanishing Lyapunov exponents, i.e., both
zero and positive Kolmogorov-Sinai entropy), for {\it finite} entropy
production per unit time \cite{classicalchaos}, whereas the others do not.

Now, Wilk and Wlodarczyk \cite{wilk} as well as Beck \cite{turbulencebeck2},
noticed that Eq. (4) can be seen as the result of complex dynamics in
stationary states associated with regions that exhibit spatio-temporal
fluctuations of temperature. Such interpretation relies on the fact that $%
e_q^{-\beta_q^\prime E_i}=\int_0^\infty d\beta \,e^{-\beta E_i} f_q(\beta)$,
where $f_q(\beta)$ is a $\chi^2$ distribution, the entropic index $q$ being
given by $q=\langle\beta^2 \rangle / \langle \beta \rangle^2$. Clearly, $%
\lim_{q\to 1} f_q(\beta)=\delta(\beta-\beta_1^\prime)$, where $\delta$
represents Dirac's delta.

Very recently, Beck and Cohen \cite{beckcohen} made a further step along
this line, which they called {\it superstatistics} (``statistics of
statistics"). They introduced a generalized BG factor, namely 
\begin{equation}
B(E)=\int_0^\infty d\beta \,e^{-\beta E} f(\beta) \;,
\end{equation}
where $f(\beta)$ is a generic admissible probability distribution of the
stochastic inverse ``temperature" satisfying $\int_0^\infty d\beta
\,f(\beta)=1$, and 
\begin{equation}
q_{BC} \equiv \langle\beta^2 \rangle / \langle \beta \rangle^2 \;,
\end{equation}
where $\langle ... \rangle \equiv \int_0^\infty d\beta (...) f(\beta)$. We
have slightly modified their original notation, by adding the subindex $BC$
(which stands for ``Beck-Cohen"), in order to avoid confusion with $q$. If $%
f(\beta)$ is a $\chi^2$ (or gamma) distribution, then $q_{BC}=q$; otherwise, those are
generically different quantities.

The reason for considering the generalization (5) of the $q$-exponential
weight appearing in nonextensive statistical mechanics basically is that,
for systems with sufficiently complex dynamics, there is no {\em a priori }
reason for not expecting the existence in nature of even more general
distributions. The effective statistical mechanics of such systems will
depend on the statistical properties of the fluctuations of the temperature,
or of even other intensive quantities. Naturally, if there are no
fluctuations of intensive quantities at all, the system must obey BG
statistical mechanics (i.e., $q_{BC}=q=1$). Beck and Cohen also showed that,
for small variance of the fluctuations (i.e., $|q_{BC}-1| <<1$), the first
order correction to the BG statistics is the nonextensive statistics \cite
{tsallis}. That is, the particular mechanism where nonextensivity is driven
by relative small fluctuations of temperature is {\it universal} and
precisely corresponds to nonextensive statistical mechanics with $q = q_{BC}$%
. The whole idea was also illustrated with the uniform, bimodal, log-normal
and $F-$ distributions.

So far so good. But it is important to realize that having a generalized
statistical weight such as (5) is {\it not} enough for having a statistical
mechanics. Indeed, consistent expressions for the entropy $S$ and for the
internal energy $U$ are also needed. By ``consistent" we mean that the
optimization of $S$ with the norm and the energy constraints should yield
Eq. (5). We have recently bridged this gap \cite{tsallisouza} and proposed
the entropic functional $S$, as well as the associated energy constraint, so
that superstatistics can be now {\it derived} from a variational principle
within a statistical mechanical frame. Let us recall that any monotonic
function of the entropic functional $S$ introduced in \cite{tsallisouza}
will also lead to Eq. (5). The entropy $S$ seems, however, to be the {\it %
physical} one. Indeed, not only it is concave by construction \cite
{tsallisouza}, but it is also stable. The proof of the latter is the main
result of the present paper.

The entropic form $S$ and the associated constraint for the internal energy $%
U$ introduced in \cite{tsallisouza} are given by (from now on $k=1$ for
simplicity) 
\begin{equation}
S=\sum_{i=1}^Ws(p_i)\;\;\;\;(s(p_i)\ge 0;\;s(0)=s(1)=0)\;,  \label{S}
\end{equation}
and 
\begin{equation}
\frac{\sum_{i=1}^Wu(p_i)E_i}{\sum_{i=1}^Wu(p_i)}=U\;\;\;\;(0\le u(p_i)\le
1;\;u(0)=0;\;u(1)=1)\;.  \label{U}
\end{equation}
For example, for the BG statistics, we have $s_{BG}(p)=-p\ln p\;$ and $%
u_{BG}(p)=p\;$, and, for the nonextensive statistics, we have $%
s_q(p)=(p-p^{\,q})/(q-1)$ and $u_q(p)=p^{\,q}$. The function $s(p)$ should
generically have a definite concavity $\forall p\in [0,1]$. Conditions (\ref
{S}) imply that $S\ge 0$ and that certainty corresponds to $S=0$. The
function $u(p)$ should generically be a monotonically increasing one.
Certainty about $E_j$ implies $U=E_j$. The quantity $u(p_i)/%
\sum_{j=1}^Wu(p_j)$ constitutes itself a probability distribution (the {\it %
escort distribution} \cite{beckbook}). Starting from the distribution of $%
f(\beta )$, we calculate $B(E)$ and $\int_0^\infty dE^{\prime }B(E^{\prime })
$, and we find the inverse function, noted $E(p)$, of $p(E)\equiv
B(E)/\int_0^\infty dE^{\prime }B(E^{\prime })$. Now, two cases are possible.
The first one corresponds to distributions where the lowest admissible value
of $E(p)$ is $E^{*}\to -\infty $. In this case\cite{tsallisouza}, $u(p)=p$
and 
\begin{equation}
s(p)=\alpha p+\int_0^p dx E(x) \;,  \label{f1}
\end{equation}
where $\alpha =-\int_0^1dx E(x)$. The second case corresponds to a finite
and known value of $E^{*}$. In this case, the final form of $s(p)$ is given
by \cite{tsallisouza} 
\begin{equation}
s(p)=E^{*}[p+u(p)],  \label{f2}
\end{equation}
where 
\begin{equation}
u(p)=\frac{\int_0^p\frac{dx}{1-E(x)/E^{*}}\;}{\int_0^1\frac{dx}{1-E(x)/E^{*}}%
}\;.  \label{f3}
\end{equation}
Eqs. (\ref{f1}), (\ref{f2}) and (\ref{f3}) completely determine the
formulation of the statistical mechanics associated with the Beck-Cohen
superstatistics.

\smallskip For a statistical quantity $O(p)$ to be an observable, or
physical quantity, a necessary condition \cite{lesche} is that, under
arbitrary small variations of the probabilities $p$, its relative variation
remains small. We then say that $O(p)$ is {\it stable}. To be more precise,
if we consider two probability sets $p$ and $p^\prime$ associated with $W$
microstates, the measure of the size of the deformation can be defined as
follows \cite{lesche} 
\begin{equation}
||p-p^{\prime }||=\sum_{i=1}^W|p_i-p_i^{\prime }| \;.  \label{dis}
\end{equation}
The condition of stability of $O(p)$ is then given by 
\begin{equation}
||p-p^{\prime }||<\delta _\epsilon \quad \Rightarrow \quad 
%TCIMACRO{\QOVERD| | {O(p)-O(p^{\prime })}{O_{\max }} }
%BeginExpansion
{O(p)-O(p^{\prime }) \overwithdelims|| O_{\max }}%
%EndExpansion
<\epsilon  \label{est}
\end{equation}
for any $\epsilon >0$, with $\delta _\epsilon >0$ being independent from $W$%
. This implies, in particular, that $\lim_{\epsilon \to 0} \lim_{W\to\infty}
R = \lim_{W\to\infty} \lim_{\epsilon \to 0} R =0$.

Using conditions (12) and (13), Lesche\cite{lesche} showed that the
R\'{e}nyi entropies are unstable, and, therefore, cannot represent
experimentally observable quantities (except, of course, for $q=1$, when
they recover the BG entropy, which is stable). Recently, using this
formalism, Abe proved\cite{abe} that $S_q$ is stable and can consequently
provide the physical entropic basis for the power-law distributions
frequently observed in nature (Abe also pointed that another candidate,
namely the so called normalized nonextensive entropies, are unstable, like
the Renyi ones). It is along these lines that we shall now prove that $S$,
as given by Eqs. (\ref{f1}), (\ref{f2}) and (\ref{f3}), is stable.

We first remark that $B(E)$ is a physical generalization of the $q$
-exponential factor (hence of the BG factor), and must therefore be a
monotonically decreasing function; the same holds then for $p(E)$ and for $%
E(p)$. It immediately follows from Eq. (\ref{f1}) that 
\begin{equation}
S(p)-S(p^{\prime })=\sum_{i=1}^W\int_{p_i^{\prime }}^{p_i^{}}dxE(x)\;.
\label{14}
\end{equation}
Notice that, since $S(p)$ is concave, we can write $S(\lambda p+(1-\lambda
)p^{\prime })\geq \lambda S(p)+(1-\lambda )S(p^{\prime })$ ($\forall \,0\leq
\lambda \leq 1$). Since $E(p)$ is positive and a monotonically decreasing
function, it follows from Eq. (\ref{f1}) that $F(p)$ $\equiv \int_0^pdxE(x)$
is a concave monotonically increasing function. Using the monotonicity of $%
E(x)$, if $p_i^{}\geq p_i^{\prime }\geq 0$, we have that 
\begin{equation}
\int_{p_i^{\prime }}^{p_i^{}}dxE(x)=\int_0^{p_i^{}-p_i^{\prime
}}dxE(x+p_i^{\prime })\leq \int_0^{p_i^{}-p_i^{\prime
}}dxE(x)=F(p_i^{}-p_i^{\prime }),
\end{equation}
and, if $p_i^{\prime }\geq p_i^{}\geq 0$, we then have that 
\begin{equation}
\int_{p_i^{\prime }}^{p_i^{}}dxE(x)\leq -\int_0^{p_i^{\prime
}-p_i^{}}dxE(x)=-F(p_i^{\prime }-p_i^{}).
\end{equation}
Considering the upperbound, we can write that 
\begin{equation}
\left| S(p)-S(p^{\prime })\right| \leq \sum_{i=1}^W\left| \int_{p_i^{\prime
}}^{p_i^{}}dxE(x)\right| \leq \sum_{i=1}^WF(|p_i^{}-p_i^{\prime }|).
\label{ine1}
\end{equation}

On the other hand, from (\ref{f2}) we obtain 
\begin{equation}
S(p)-S(p^{\prime })=E^{*}\sum_{i=1}^W[u(p_i^{})-u(p_i^{\prime })]\;.
\end{equation}
Using that $u(p)$ is a monotonically increasing function and that $S(p)$ is
concave, we can see from Eq. (\ref{f2}) that $u(p)$ is a concave
monotonically increasing function if $E^{*}>0$, and it is a convex
monotonically increasing function if $E^{*}<0$. Hence, $F^{\prime
}(p)=E^{*}u(p)$ always is a concave monotonically increasing function. It
follows that, if $p_i^{}\geq p_i^{\prime }\geq 0$, we have $F^{\prime
}(p_i)-F^{\prime }(p_i^{\prime })\leq F^{\prime }(p_i-p_i^{\prime })\;,$and,
if $p_i^{\prime }\geq p_i\geq 0$, we have $-[F^{\prime }(p_i)-F^{\prime
}(p_i^{\prime })]\leq F^{\prime }(p_i^{\prime }-p_i)$. This implies that $%
\left| F^{\prime }(p_i)-F^{\prime }(p_i^{\prime })\right| \leq F^{\prime
}(|p_i^{}-p_i^{\prime }|)$, then
\begin{equation}
\left| S(p)-S(p^{\prime })\right| \leq \sum_{i=1}^W\left| F^{\prime
}(p_i)-F^{\prime }(p_i^{\prime })\right| \leq \sum_{i=1}^WF^{\prime
}(|p_i^{}-p_i^{\prime }|)\;.  \label{ine2}
\end{equation}

We now use that $g(\lambda x)g(y)\leq g(x)g(\lambda y)$ for any increasing
concave function $g(x)$, with $g(0)\geq 0$, $0<x<y$ and $0\leq \lambda <1$.
Considering $\lambda =||p-p^{\prime }||<1$, $x=\frac{|p_i^{}-p_i^{\prime }|}{%
||p-p^{\prime }||}$ and $y=1$ \cite{nau}, it follows that 
\begin{equation}
g(|p_i^{}-p_i^{\prime }|)\leq \frac{g(||p-p^{\prime }||)}{g(1)}\quad g\left( 
\frac{|p_i^{}-p_i^{\prime }|}{||p-p^{\prime }||}\right) .
\end{equation}
Finally, since $F(p)$ and $F^{\prime }(p)$ are increasing concave functions,
from Eqs.(\ref{ine1}) and (\ref{ine2}), if we identify $g(p)$ as being $F(p)$
or $F^{\prime }(p)$, we have 
\begin{equation}
|S(p)-S(p^{\prime })|\leq \frac{g(||p-p^{\prime }||)}{g(1)}\left[ S\left( 
\frac{|p_i^{}-p_i^{\prime }|}{||p-p^{\prime }||}\right) +g(1)\right] \;.\quad
\end{equation}
Therefore, denoting by $S_{\max }$ the maximal entropy, we obtain 
\begin{equation}
|S(p)-S(p^{\prime })|\leq \frac{g(||p-p^{\prime }||)}{g(1)}\left[ S_{\max
}+g(1)\right] \; .\quad
\end{equation}
This inequality shows that, for $\epsilon >0$ such that $|S(p)-S(p^{\prime
})|\leq \epsilon S_{\max }$, we have that %$\Rightarrow $ 
$||p-p^{\prime }||\leq \delta _\epsilon ^{}$, which proves the stability of
the entropy for superstatistics.

Summarizing, the entropies introduced in Ref. \cite{tsallisouza} have been
studied. Since they are simultaneously concave and stable, it is allowed to
think that these entropies are physical ones, and as such can be used to
formulate a proper statistical mechanics which further generalizes the
current nonextensive statistical mechanics. The high precision experimental
and numerical data for fully developed turbulence recently discussed by Beck 
\cite{beckrecent} could well be an important physical application.

Useful remarks from E. Brigatti are acknowledged, as well as partial support
from PCI/MCT, CNPq, PRONEX/MCT and FAPERJ (Brazilian agencies).


\begin{references}
\bibitem{turbulencebeck1}  C. Beck, G. S. Lewis and H. L. Swinney, Phys.
Rev. E {\bf 63}, 035303 (2001).

\bibitem{turbulencebeck2}  C. Beck, Phys. Rev. Lett. {\bf 87}, 180601 (2001).

\bibitem{turbulencearimitsu}  T. Arimitsu and N. Arimitsu, Physica A {\bf 305%
}, 218 (2002).

\bibitem{bediaga}  I. Bediaga, E.M.F. Curado and J. Miranda, Physica A {\bf %
286}, 156 (2000); C. Beck, Physica A {\bf 286}, 164 (2000).

\bibitem{rafelski}  D.B. Walton and J. Rafelski, Phys. Rev. Lett. {\bf 84},
31 (2000).

\bibitem{hydra}  A. Upadhyaya, J.-P. Rieu, J.A. Glazier and Y. Sawada,
Physica A {\bf 293}, 549 (2001).

\bibitem{classicalchaos}  M.L. Lyra and C. Tsallis, Phys. Rev. Lett. {\bf 80}%
, 53 (1998); F. Baldovin and A. Robledo, Europhys. Lett. {\bf 60}, 518
(2002); F. Baldovin and A. Robledo, Phys. Rev. E {\bf 66}, R045104 (2002);
E.P. Borges, C. Tsallis, G.F.J. Ananos and P.M.C. Oliveira, Phys. Rev. Lett. 
{\bf 89}, 254103 (2002).

\bibitem{quantumchaos}  Y.S. Weinstein, S. Lloyd and C. Tsallis, Phys. Rev.
Lett. {\bf 89}, 214101 (2002).

\bibitem{longrange}  C. Anteneodo and C. Tsallis, Phys. Rev. Lett {\bf 80},
5313 (1998); M.-C. Firpo and S. Ruffo, J. Phys. A {\bf 34}, L511 (2001); V.
Latora, A. Rapisarda and C. Tsallis, Phys. Rev. E {\bf 64}, 056134 (2001);
A. Campa, A. Giansanti and D. Moroni, Physica A {\bf 305}, 137 (2002);
B.J.C. Cabral and C. Tsallis, Phys. Rev. E {\bf 66}, 065101(R) (2002); C.
Anteneodo and R.O. Vallejos, Phys. Rev. E {\bf 65}, 016210 (2002); R.O.
Vallejos and C. Anteneodo, Phys. Rev. E {\bf 66}, 021110 (2002); M.A.
Montemurro, F. Tamarit and C. Anteneodo, Phys. Rev. E (2003), in press
[cond-mat/0205355].

\bibitem{lisa}  L. Borland, Phys. Rev. Lett. {\bf 89}, 098701 (2002).

\bibitem{reviews}  S.R.A. Salinas and C. Tsallis, eds., {\it Nonextensive
Statistical Mechanics and Thermodynamics}, Braz. J. Phys. {\bf 29}, No. 1
(1999); S. Abe and Y. Okamoto, eds., {\it Nonextensive Statistical Mechanics
and its Applications}, Series {\it Lecture Notes in Physics}
(Springer-Verlag, Berlin, 2001); G. Kaniadakis, M. Lissia and A. Rapisarda,
eds., {\it Non Extensive Statistical Mechanics and Physical Applications},
Physica A {\bf 305}, No 1/2 (Elsevier, Amsterdam, 2002); M. Gell-Mann and C.
Tsallis, eds., {\it Nonextensive Entropy - Interdisciplinary Applications}
(Oxford University Press, 2003), in preparation; H.L. Swinney and C.
Tsallis, eds., {\it Anomalous Distributions, Nonlinear Dynamics, and
Nonextensivity}, Physica D (2003), in preparation. A regularly updated
bibliography can be found at the web site
http://tsallis.cat.cbpf.br/biblio.htm .

\bibitem{tsallis}  C. Tsallis, J. Stat. Phys. {\bf 52}, 479 (1988).

\bibitem{TMP}  C. Tsallis, R.S. Mendes and A.R. Plastino,Physica A {\bf 261}%
, 534 (1998); see also E.M.F. Curado and C. Tsallis, J. Phys. A {\bf 24},
L69 (1991) [Corrigenda: {\bf 24}, 3187 (1991) and {\bf 25}, 1019 (1992)].

\bibitem{lesche}  B. Lesche, J. Stat. Phys. {\bf 27}, 419 (1982).

\bibitem{abestable}  S. Abe, Phys. Rev. E {\bf 66}, 046134 (2002).

\bibitem{santosabe}  R.J.V. dos Santos, J. Math. Phys. {\bf 38}, 4104
(1997); S. Abe, Phys. Lett. A {\bf 271}, 74 (2000).

\bibitem{wilk}  G. Wilk and Z. Wlodarczyk, Phys. Rev. Lett. {\bf 84}, 2770
(2000).

\bibitem{beckcohen}  C. Beck and E.G.D. Cohen, Physica A {\bf 321} (2003),
in press [cond-mat/0205097].

\bibitem{tsallisouza}  C. Tsallis and A.M.C. Souza, Phys. Rev. E {\bf 67},
0261XX (1 Feb 2003), in press [cond-mat/0206044].

\bibitem{beckbook}  C. Beck and F. Schlogl, {\it Thermodynamics of Chaotic
Systems} (Cambridge University Press, Cambridge, 1993).

\bibitem{abe}  S. Abe, Phys. Rev. E {\bf 66}, 046134 (2002).

\bibitem{nau}  J. Naudts, math-ph/0208038.

\bibitem{beckrecent}  C. Beck, cond-mat/0212566 (2002).
\end{references}
\end{document}